\documentclass[%
 reprint,
 10pt,
 amsmath,amssymb,
 aps,
longbibliography]{revtex4-1}

\usepackage{graphicx}
\usepackage{dcolumn}
\usepackage{bm}
\usepackage{footmisc}
\usepackage{amsmath,amssymb} 
\usepackage{mathrsfs}
\usepackage{amsfonts}
\usepackage{pgfplots}
\pgfplotsset{compat=1.5}
\usepackage{floatrow}
\usepackage{soul}
\usepackage{hyperref}

\def\IB#1{\boldsymbol{#1}} 
\def\W/!i#1{\Wi} 
\def\bten#1{\IB{\mathsf{#1}}}

\begin{document}

\preprint{APS/123-QED}

\title{Comment on ``Migration of an electrophoretic particle in a weakly inertial or viscoelastic shear flow''}

\author{Akash Choudhary\textsuperscript}
\author{T. Renganathan\textsuperscript}%
\author{S. Pushpavanam\textsuperscript}
\homepage{spush@iitm.ac.in}
 
\affiliation{%
Indian Institute of Technology Madras, Chennai, 600036 TN, India
}%

\begin{abstract}
	A recent article (\textit{Khair and Kabarowski; Phys. Rev. Fluids 5, 033702}) has studied the cross-streamline migration of electrophoretic particles in unbounded shear flows with weak inertia or viscoelasticity. That work compares their results with those reported in two of our previous studies (Choudhary et al. \textit{J. Fluid Mech. 874; J. Fluid Mech. 898}) and reports a disagreement in the derived analytical expressions. In this comment, we resolve this discrepancy. For viscoelastic flows, we show that Khair and Kabarowski  have not accounted for a leading order surface integral of polymeric stress in their calculation of first-order viscoelastic lift. When this integral is included, the resulting migration velocity matches exactly with that reported in our work (\textit{J. Fluid Mech. 898}). 
	This qualitatively changes migration direction that is reported by Khair and Kabarowski for viscoelastic flows.
	\textcolor{black}{For inertial flows, we clarify that Khair and Kabarowski find the coefficient of lift to be 1.75$\pi$, compared to 2.35$ \pi $ in our previous work (\textit{J. Fluid Mech. 874}). We show that this difference occurs because Khair and Kabarowski accurately include the effect of rapidly decaying $ \sim O(1/r^{4}) $ velocity field (a correction to the stresslet field $ \sim 1/r^{2} $), which was neglected in our previous work (\textit{J. Fluid Mech. 874}).}
\end{abstract}

\maketitle


 Ref.\cite{khair2020migration} used perturbation theory to find the lift force on an electrophoretic particle in (i) weakly inertial and (ii) weakly viscoelastic unbounded shear flows. 
The field variables were perturbed in Reynolds and Weissenberg numbers associated with both shear and electrophoresis (eq. 12-13 therein).
The reciprocal theorem is used to derive leading order lift for cases (i) \& (ii). 
They obtain the following equation (eq. 20 in ref.\cite{khair2020migration}) and use it for both cases:
\begin{equation}\label{eq1}
\IB{F}_1 = \int_{S_{p}} \IB{\sigma}_{1}\cdot \IB{n} dS = - \int_{V} \bten{H}_{test} \cdot \IB{f} dV,
\end{equation}
where $ \bten{H} $ is the test field (or auxiliary field) tensor, representing the stokeslet + source-dipole fields associated with the cross-stream motion.

\textcolor{black}{Ref.\cite{khair2020migration} evaluates the cross-stream migration for particle undergoing electrophoresis at an arbitrary angle to shear flows in an unbounded domain.
	For electric fields applied parallel to the flow, ref.\cite{khair2020migration} drew comparisons with our previous investigations for parallel orientation \cite{choudhary2020electrokinetically,choudhary2019inertial}; our work takes into account the flow curvature, wall-induced hydrodynamic and electrostatic effects at the leading order.
	They reported the migration to be identical in scaling, but different in coefficients. For viscoelastic flows, their migration points in the direction opposite to that reported in one of our previous study \cite{choudhary2020electrokinetically}.}

\begin{table}[htbp]
	\centering
	\begin{tabular}{ |c|c|c| } 
		\hline
	\textit{Velocity/Lift}  & \textit{Ref.\cite{khair2020migration}}   & \textit{Ref.\cite{choudhary2019inertial,choudhary2020electrokinetically}}\\
	\hline
	$ U_{visc}^{mig}/(\Psi_{1}^{*} U_{ep}^{*}  \, \dot{\gamma}^{*} / {\mu^{*}}) $   & $ \frac{3}{8} \left(1-\frac{10 \Psi_{2}^{*}}{3 \Psi_{1}^{*}}\right)$  & $ \frac{-1}{32 \pi}\left( 1+ \frac{2 \Psi_{2}^{*}}{\Psi_{1}^{*}} \right) $ \\
	$ F_{inertia}^{mig}/(Re \, U_{ep}^{*}  \mu^{*} a^{*}) $ &  $ \frac{7 \pi}{4} $   & $ \frac{47}{80} $  \\ 
		\hline
	\end{tabular}
		\caption{Comparison of migration velocity and lift force coefficients reported by ref.\cite{khair2020migration}. Here, $ \Psi_{1}^{*} $ \& $ \Psi_{2}^{*} $ are the first \& second normal stress coefficients, and $ U_{visc}^{mig} $ \& $ F_{inertia}^{mig} $  represent the viscoelastic migration velocity and inertial lift force, respectively; $ U_{ep}^{*} $ is the dimensional electrophoretic velocity ($ \epsilon_{m}^{*} \zeta_{p}^{*} E_{\infty}^{*}/\mu^{*} $); $ \dot{\gamma}^{*} $ is the dimensional shear in the background flow; $ Re $ represents the Reynolds number ($ \rho^{*} \dot{\gamma}^{*} a^{*}/\mu^{*} $), where $ a^{*} $ is the dimensional particle radius and $ \rho^{*} $ is the fluid density.}
\label{table1}
\end{table}

The comparison of coefficients reported by ref.\cite{khair2020migration} is described in table.\ref{table1}.
They converted the coefficients in ref.\cite{choudhary2019inertial,choudhary2020electrokinetically} to compare with theirs. We shall show that a factor of $ 4\pi $ has to be multiplied to ref.\cite{choudhary2019inertial,choudhary2020electrokinetically} for an equitable comparison.\\

\textit{A factor of $ 4\pi $}:
Following the classical works of Anderson and co-workers \cite{anderson1989,keh1985}, in our previous works \cite{choudhary2019inertial,choudhary2020electrokinetically}, we had taken the electrophoretic velocity as $ \epsilon^{*} \zeta_{p}^{*}  E_{\infty}^{*}/(4\pi \mu^{*} ) $, where $ \epsilon^{*}, \zeta_{p}^{*}, \phi^{*}, \mu^{*} $ are dielectric constant, particle zeta potential, electrostatic potential and viscosity, respectively; $ * $ represents dimensional variables. Ref.\cite{khair2020migration} have absorbed the factor $ 4\pi $ inside the dielectric constant $ \epsilon^{*} $. 
Thus, for an appropriate comparison, the conversions in table-\ref{table1} (of our results in ref.\cite{choudhary2019inertial,choudhary2020electrokinetically}) must be multiplied with $ 4\pi $ \footnote{While comparing the results, the Hartmann number ($ Ha $) that should be accounted is $ \epsilon^{*}_{m} E_{\infty}^{*\,2} a^{*}/ \mu^{*} U_{ch} $ rather than $ \epsilon^{*} E_{\infty}^{*\,2} a^{*}/4 \pi \mu^{*} U_{ch} $ ($ U_{ch} $ is the characteristic velocity scale), where $ \epsilon_{m}^{*} = \epsilon^{*}/4\pi $.}.
The equitable comparison of results is shown in table-\ref{tab2}, which will be referred to address the differences.

\begin{table}[htbp]
	\centering
	\begin{tabular}{ |c|c|c| } 
		\hline
		\textit{Velocity/Lift} & \textit{Ref.\cite{khair2020migration}}  &		\textit{Ref.\cite{choudhary2019inertial,choudhary2020electrokinetically}}\\ 
		\hline 
		$ U_{visc}^{mig}/(\Psi_{1}^{*} U_{ep}^{*}  \, \dot{\gamma}^{*} / {\mu^{*}}) $   & $ \frac{3}{8} \left(1-\frac{10 \Psi_{2}^{*}}{3 \Psi_{1}^{*}}\right)$  & $ \frac{-1}{8}\left( 1+ \frac{2 \Psi_{2}^{*}}{\Psi_{1}^{*}} \right) $ \\
		$ F_{inertia}^{mig}/(Re \, U_{ep}^{*}  \mu^{*} a^{*}) $  &  $ \frac{7 \pi}{4} $   & $ \frac{47\pi}{20} $ \\ 
		\hline
	\end{tabular}
		\caption{Equitable comparison of migration velocity and lift force coefficients: a factor of $ 4\pi $ is multiplied to the coefficients of ref.\cite{choudhary2019inertial,choudhary2020electrokinetically}.}
\label{tab2}
\end{table}

\textit{Viscoelasticity:}
Ref.\cite{khair2020migration} used the same derivation for both the inertial and viscoelastic migration: a body force acts on the particle which is captured by 
$ - \int_{V} \bten{H}_{test} \cdot \IB{f} dV $. They derive it for the case of inertia and then use that result directly for the case of viscoelastic fluid (see eq.20 therein).
This very extension leaves out an important contribution because, in viscoelastic fluids, eq.(\ref{eq1}) does not represent the complete first-order force on a spherical body.
An extra surface integral of leading order polymeric stress must be added.

This can be easily seen through a perturbation expansion of the total stress tensor in Weissenberg number (denoted as Deborah number in ref.\cite{choudhary2020electrokinetically}). The non-dimensional stress tensor for a second-order-fluid is:
\begin{equation}
\bten{T} = -  p \bten{I}  + \bten{e} + W\!i \, \bten{\sigma}^{P},
\end{equation}
where, $ p $ represents pressure, $ \bten{e} $ represents the rate-of-strain tensor,  $ \bten{\sigma}^{P} $ is the polymeric stress. The momentum equation is governed by $ \nabla \cdot \bten{T } = 0 $: 
\begin{equation}
\nabla \cdot (-  p \bten{I}  + \bten{e}) = - W\!i \, \nabla \cdot \bten{\sigma}^{P}
\end{equation}
A perturbation expansion of velocity and pressure field in Weissenberg would yield the first order problem (i.e. $ O(Wi) $) as 
\begin{equation}\label{firstOrder}
\nabla^{2} \IB{u}_{1} - \nabla p_{1} = - \nabla \cdot \bten{\sigma}^{P}_{0} \equiv \IB{f}_{0}
\end{equation}
This is a non-Newtonian equivalent to eq.14 in ref.\cite{khair2020migration}, where the right hand side is the `{viscoelastic body force}' $ \IB{f}_{0} $, which is known, provided the creeping flow problem is known. 
We use (\ref{firstOrder}) and perform the steps outlined in ref.\cite{khair2020migration}, and arrive at:
\begin{equation}\label{mis}
\int_{S} (-  p_{1} \bten{I}  + \bten{e}_{1}) \cdot \IB{n} \; dS = - \int_{V} \bten{H}_{test} \cdot \IB{f}_{0} \; dV.
\end{equation}
The left hand side of the above expression accurately represents the first order (inertial) force for Newtonian fluids \cite{ho1974,kim2013}.
Ref.\cite{khair2020migration} derive (\ref{mis}) and evaluate it to find the lift for an inertial shear flow. 
The same equation is also used to evaluate lift for viscoelastic shear flow, thereby leaving out an important contribution.
This is because, for a second-order fluid, the left hand side of (\ref{mis}) does not amount to the total first order force; a surface integral of leading order polymeric stress is missing. 
A correct expression for the first order viscoelastic force is obtained by adding surface integral of the leading order polymeric stress on both sides of (\ref{mis}).
\begin{eqnarray}\label{F1}
\int_{S} (-  p_{1} \bten{I}  + \bten{e}_{1} +   \bten{\sigma}^{P}_{0}) \cdot \IB{n}  dS &=& - \int_{V} \bten{H}_{test} \cdot \IB{f}_{0}  dV  \nonumber \\
& &+ { {\; \int_{S}   \, \bten{\sigma}^{P}_{0} \cdot \IB{n} \, dS}} .
\end{eqnarray}
In the above equation, the left hand side is the total first order viscoelastic force ($ \IB{F}_{1} $). The right hand side is equivalent to the widely used viscoelastic bulk force expression at the leading order: $ -\int_{V} \IB{\sigma}^{P}_{0}:\nabla\IB{u}_{test} \,dV $ \cite{choudhary2020electrokinetically,corato2015locomotion,datt2017activeComplex,ho1976migration,elfring2015theory}.
The importance of addition of the surface integral, when drawing parallels between inertial and viscoelastic force, (and the equivalence of two expressions) is also discussed in the past by the pioneering works of Leal \cite[p.314]{leal1975slow}\cite[p.790]{ho1976migration}.
\\

The contribution of this additional surface integral to the migration velocity is \footnote{A mathematica code for the evaluation of this surface integral is provided in the supplementary material.}:
\begin{eqnarray}
U^{mig}_{extra} &=& \left( \frac{-\Psi_{1}^{*}}{2} + \Psi_{2}^{*}  \right) \frac{U_{ep}^{*}  \, \dot{\gamma}^{*}}{\mu^{*}}.
\end{eqnarray}
Adding the above component to that reported by ref.\cite{khair2020migration}, we obtain the migration velocity as:
\begin{eqnarray}
U^{mig}_{total} &=& \frac{-1}{8} \left( 1 + \frac{2 \Psi_{2}^{*}}{\Psi_{1}^{*}} \right) \Psi_{1}^{*} \frac{U_{ep}^{*}  \, \dot{\gamma}^{*}}{\mu^{*}}.
\end{eqnarray}
This coefficient is identical to that reported by \cite{choudhary2020electrokinetically} (see table-\ref{tab2}). This reverses the migration direction predicted by ref.\cite{khair2020migration}.
\\

In their appendix section, ref.\cite{khair2020migration} use the special case of $ \Psi_{2}^{*}/\Psi_{1}^{*}  = -1/2 $ to independently verify their results.
For this special case, the Giesekus theorem states that only the pressure field is perturbed at the leading order; first order perturbed velocity field is zero i.e. $ \IB{v}_{1}=0 $, yielding $ \bten{e}_{1}=0 $. From eq.A3 of ref.\cite{khair2020migration}, it can be seen that the first order viscoelastic force is taken to be $ \int_{S} -p_{1} \bten{I}  \cdot \IB{n} \, dS $.
However, as explained above, the total first order force on the particle should be:
\textsl{}$ \int_{S} (-p_{1} \bten{I} + \bten{\sigma}^{P}_{0} ) \cdot \IB{n} \; dS, $
because the leading order stress tensor has an additional polymeric stress. Specifically, this part is the corotational component of the polymeric stress in the limit $ \Psi_{2}^{*}/\Psi_{1}^{*}  = -1/2 $ \cite{koch2006stress}.
Our previous work \cite{choudhary2020electrokinetically} uses the Giesekus theorem and finds the viscoelastic migration for this special case (see Appendix D.1 ref.\cite{choudhary2020electrokinetically}). Therein, we show the contribution to lift from pressure and corotational stress, and find that the pressure contribution is identical to that reported in ref.\cite[eq. A3]{khair2020migration}\footnote{The coefficient is $ 6\pi \Psi_{1}^{*} $.}.
\\



\textcolor{black}{\textit{Inertia}:
	As shown in table-\ref{tab2}, the comparison of coefficients is essentially: $ 1.75 \pi $ (in ref.\cite{khair2020migration}) and $ 2.35\pi $ (in our work \cite{choudhary2019inertial}). This difference occurs because ref.\cite{khair2020migration}, precisely, includes the effect of rapidly decaying velocity field ($ \sim 1/r^{4} $); a correction to the stresslet field $ (\sim 1/r^{2}) $. In our formulation (ref.\cite{choudhary2019inertial}), we had not accounted for it because the aim was to include the effects of slowly decaying fields and their wall-reflections.
	To confirm this speculation, we incorporate the $ O(1/r^{4}) $ velocity disturbance in the inner integral of our formulation in ref.\cite[eq. 4.7]{choudhary2019inertial}. Upon integration, an exact match is obtained with their coefficient for unbounded shear flows (i.e. $ 1.75 \pi $). }

\pagebreak

\textcolor{black}{\textit{Conclusions:} 
	In this comment, we show that for viscoelastic migration, ref.\cite{khair2020migration} have missed a surface integral of leading order polymeric stress in (i) their derivation for second-order fluid and (ii) for the special case of $ \Psi_{2}/\Psi_{1} = -1/2 $. When this is included, the migration velocity matches exactly with our previous work \cite{choudhary2020electrokinetically}; refer supplementary material for a mathematica code that details the evaluation. This qualitatively changes migration direction that is reported by ref.\cite{khair2020migration}.
	For the case of inertial migration in unbounded domains, we clarify that the results of ref.\cite{khair2020migration} are more accurate because they include the effect of a rapidly decaying $ O(1/r^{4}) $ field, which was neglected in our previous work \cite{choudhary2019inertial}. We also provide a mathematica code (see supplementary material) which details how this inclusion corrects the coefficient of lift force from 2.35$ \pi $ (corresponding to our work \cite{choudhary2019inertial}) to 1.75$ \pi $ (corresponding to ref.\cite{khair2020migration}). 
	This change in coefficient does not qualitatively change the results for inertial lift force in our previous work \cite{choudhary2019inertial}.}


\bibliography{Akash}

\begin{thebibliography}{16}%
\makeatletter
\providecommand \@ifxundefined [1]{%
 \@ifx{#1\undefined}
}%
\providecommand \@ifnum [1]{%
 \ifnum #1\expandafter \@firstoftwo
 \else \expandafter \@secondoftwo
 \fi
}%
\providecommand \@ifx [1]{%
 \ifx #1\expandafter \@firstoftwo
 \else \expandafter \@secondoftwo
 \fi
}%
\providecommand \natexlab [1]{#1}%
\providecommand \enquote  [1]{``#1''}%
\providecommand \bibnamefont  [1]{#1}%
\providecommand \bibfnamefont [1]{#1}%
\providecommand \citenamefont [1]{#1}%
\providecommand \href@noop [0]{\@secondoftwo}%
\providecommand \href [0]{\begingroup \@sanitize@url \@href}%
\providecommand \@href[1]{\@@startlink{#1}\@@href}%
\providecommand \@@href[1]{\endgroup#1\@@endlink}%
\providecommand \@sanitize@url [0]{\catcode `\\12\catcode `\$12\catcode
  `\&12\catcode `\#12\catcode `\^12\catcode `\_12\catcode `\%12\relax}%
\providecommand \@@startlink[1]{}%
\providecommand \@@endlink[0]{}%
\providecommand \url  [0]{\begingroup\@sanitize@url \@url }%
\providecommand \@url [1]{\endgroup\@href {#1}{\urlprefix }}%
\providecommand \urlprefix  [0]{URL }%
\providecommand \Eprint [0]{\href }%
\providecommand \doibase [0]{http://dx.doi.org/}%
\providecommand \selectlanguage [0]{\@gobble}%
\providecommand \bibinfo  [0]{\@secondoftwo}%
\providecommand \bibfield  [0]{\@secondoftwo}%
\providecommand \translation [1]{[#1]}%
\providecommand \BibitemOpen [0]{}%
\providecommand \bibitemStop [0]{}%
\providecommand \bibitemNoStop [0]{.\EOS\space}%
\providecommand \EOS [0]{\spacefactor3000\relax}%
\providecommand \BibitemShut  [1]{\csname bibitem#1\endcsname}%
\let\auto@bib@innerbib\@empty
\bibitem [{\citenamefont {Khair}\ and\ \citenamefont
  {Kabarowski}(2020)}]{khair2020migration}%
  \BibitemOpen
  \bibfield  {author} {\bibinfo {author} {\bibfnamefont {Aditya~S}\
  \bibnamefont {Khair}}\ and\ \bibinfo {author} {\bibfnamefont {Jason~K}\
  \bibnamefont {Kabarowski}},\ }\bibfield  {title} {\enquote {\bibinfo {title}
  {Migration of an electrophoretic particle in a weakly inertial or
  viscoelastic shear flow},}\ }\href@noop {} {\bibfield  {journal} {\bibinfo
  {journal} {Physical Review Fluids}\ }\textbf {\bibinfo {volume} {5}},\
  \bibinfo {pages} {033702} (\bibinfo {year} {2020})}\BibitemShut {NoStop}%
\bibitem [{\citenamefont {Choudhary}\ \emph {et~al.}(2020)\citenamefont
  {Choudhary}, \citenamefont {Li}, \citenamefont {Renganathan}, \citenamefont
  {Xuan},\ and\ \citenamefont {Pushpavanam}}]{choudhary2020electrokinetically}%
  \BibitemOpen
  \bibfield  {author} {\bibinfo {author} {\bibfnamefont {Akash}\ \bibnamefont
  {Choudhary}}, \bibinfo {author} {\bibfnamefont {Di}~\bibnamefont {Li}},
  \bibinfo {author} {\bibfnamefont {T}~\bibnamefont {Renganathan}}, \bibinfo
  {author} {\bibfnamefont {Xiangchun}\ \bibnamefont {Xuan}}, \ and\ \bibinfo
  {author} {\bibfnamefont {S}~\bibnamefont {Pushpavanam}},\ }\bibfield  {title}
  {\enquote {\bibinfo {title} {Electrokinetically enhanced cross-stream
  particle migration in viscoelastic flows},}\ }\href@noop {} {\bibfield
  {journal} {\bibinfo  {journal} {Journal of Fluid Mechanics}\ }\textbf
  {\bibinfo {volume} {898}} (\bibinfo {year} {2020})}\BibitemShut {NoStop}%
\bibitem [{\citenamefont {Choudhary}\ \emph {et~al.}(2019)\citenamefont
  {Choudhary}, \citenamefont {Renganathan},\ and\ \citenamefont
  {Pushpavanam}}]{choudhary2019inertial}%
  \BibitemOpen
  \bibfield  {author} {\bibinfo {author} {\bibfnamefont {A.}~\bibnamefont
  {Choudhary}}, \bibinfo {author} {\bibfnamefont {T.}~\bibnamefont
  {Renganathan}}, \ and\ \bibinfo {author} {\bibfnamefont {S.}~\bibnamefont
  {Pushpavanam}},\ }\bibfield  {title} {\enquote {\bibinfo {title} {Inertial
  migration of an electrophoretic rigid sphere in a two-dimensional poiseuille
  flow},}\ }\href {\doibase 10.1017/jfm.2019.479} {\bibfield  {journal}
  {\bibinfo  {journal} {Journal of Fluid Mechanics}\ }\textbf {\bibinfo
  {volume} {874}},\ \bibinfo {pages} {856--890} (\bibinfo {year}
  {2019})}\BibitemShut {NoStop}%
\bibitem [{\citenamefont {Anderson}(1989)}]{anderson1989}%
  \BibitemOpen
  \bibfield  {author} {\bibinfo {author} {\bibfnamefont {John~L}\ \bibnamefont
  {Anderson}},\ }\bibfield  {title} {\enquote {\bibinfo {title} {Colloid
  transport by interfacial forces},}\ }\href@noop {} {\bibfield  {journal}
  {\bibinfo  {journal} {Annual review of fluid mechanics}\ }\textbf {\bibinfo
  {volume} {21}},\ \bibinfo {pages} {61--99} (\bibinfo {year}
  {1989})}\BibitemShut {NoStop}%
\bibitem [{\citenamefont {Keh}\ and\ \citenamefont {Anderson}(1985)}]{keh1985}%
  \BibitemOpen
  \bibfield  {author} {\bibinfo {author} {\bibfnamefont {Huan-Jang}\
  \bibnamefont {Keh}}\ and\ \bibinfo {author} {\bibfnamefont {JL}~\bibnamefont
  {Anderson}},\ }\bibfield  {title} {\enquote {\bibinfo {title} {Boundary
  effects on electrophoretic motion of colloidal spheres},}\ }\href@noop {}
  {\bibfield  {journal} {\bibinfo  {journal} {Journal of Fluid Mechanics}\
  }\textbf {\bibinfo {volume} {153}},\ \bibinfo {pages} {417--439} (\bibinfo
  {year} {1985})}\BibitemShut {NoStop}%
\bibitem [{Note1()}]{Note1}%
  \BibitemOpen
  \bibinfo {note} {While comparing the results, the Hartmann number ($ Ha $)
  that should be accounted is $ \epsilon ^{*}_{m} E_{\infty }^{*\protect
  \tmspace +\thinmuskip {.1667em}2} a^{*}/ \mu ^{*} U_{ch} $ rather than $
  \epsilon ^{*} E_{\infty }^{*\protect \tmspace +\thinmuskip {.1667em}2}
  a^{*}/4 \pi \mu ^{*} U_{ch} $ ($ U_{ch} $ is the characteristic velocity
  scale), where $ \epsilon _{m}^{*} = \epsilon ^{*}/4\pi $.}\BibitemShut
  {Stop}%
\bibitem [{\citenamefont {Ho}\ and\ \citenamefont {Leal}(1974)}]{ho1974}%
  \BibitemOpen
  \bibfield  {author} {\bibinfo {author} {\bibfnamefont {BP}~\bibnamefont
  {Ho}}\ and\ \bibinfo {author} {\bibfnamefont {LG}~\bibnamefont {Leal}},\
  }\bibfield  {title} {\enquote {\bibinfo {title} {Inertial migration of rigid
  spheres in two-dimensional unidirectional flows},}\ }\href@noop {} {\bibfield
   {journal} {\bibinfo  {journal} {Journal of fluid mechanics}\ }\textbf
  {\bibinfo {volume} {65}},\ \bibinfo {pages} {365--400} (\bibinfo {year}
  {1974})}\BibitemShut {NoStop}%
\bibitem [{\citenamefont {Kim}\ and\ \citenamefont {Karrila}(2013)}]{kim2013}%
  \BibitemOpen
  \bibfield  {author} {\bibinfo {author} {\bibfnamefont {Sangtae}\ \bibnamefont
  {Kim}}\ and\ \bibinfo {author} {\bibfnamefont {Seppo~J}\ \bibnamefont
  {Karrila}},\ }\href@noop {} {\emph {\bibinfo {title} {Microhydrodynamics:
  principles and selected applications}}}\ (\bibinfo  {publisher} {Courier
  Corporation},\ \bibinfo {year} {2013})\BibitemShut {NoStop}%
\bibitem [{\citenamefont {De~Corato}\ \emph {et~al.}(2015)\citenamefont
  {De~Corato}, \citenamefont {Greco},\ and\ \citenamefont
  {Maffettone}}]{corato2015locomotion}%
  \BibitemOpen
  \bibfield  {author} {\bibinfo {author} {\bibfnamefont {M}~\bibnamefont
  {De~Corato}}, \bibinfo {author} {\bibfnamefont {F}~\bibnamefont {Greco}}, \
  and\ \bibinfo {author} {\bibfnamefont {PL}~\bibnamefont {Maffettone}},\
  }\bibfield  {title} {\enquote {\bibinfo {title} {Locomotion of a
  microorganism in weakly viscoelastic liquids},}\ }\href@noop {} {\bibfield
  {journal} {\bibinfo  {journal} {Physical Review E}\ }\textbf {\bibinfo
  {volume} {92}},\ \bibinfo {pages} {053008} (\bibinfo {year}
  {2015})}\BibitemShut {NoStop}%
\bibitem [{\citenamefont {Datt}\ \emph {et~al.}(2017)\citenamefont {Datt},
  \citenamefont {Natale}, \citenamefont {Hatzikiriakos},\ and\ \citenamefont
  {Elfring}}]{datt2017activeComplex}%
  \BibitemOpen
  \bibfield  {author} {\bibinfo {author} {\bibfnamefont {Charu}\ \bibnamefont
  {Datt}}, \bibinfo {author} {\bibfnamefont {Giovanniantonio}\ \bibnamefont
  {Natale}}, \bibinfo {author} {\bibfnamefont {Savvas~G}\ \bibnamefont
  {Hatzikiriakos}}, \ and\ \bibinfo {author} {\bibfnamefont {Gwynn~J}\
  \bibnamefont {Elfring}},\ }\bibfield  {title} {\enquote {\bibinfo {title} {An
  active particle in a complex fluid},}\ }\href@noop {} {\bibfield  {journal}
  {\bibinfo  {journal} {Journal of Fluid Mechanics}\ }\textbf {\bibinfo
  {volume} {823}},\ \bibinfo {pages} {675--688} (\bibinfo {year}
  {2017})}\BibitemShut {NoStop}%
\bibitem [{\citenamefont {Ho}\ and\ \citenamefont
  {Leal}(1976)}]{ho1976migration}%
  \BibitemOpen
  \bibfield  {author} {\bibinfo {author} {\bibfnamefont {BP}~\bibnamefont
  {Ho}}\ and\ \bibinfo {author} {\bibfnamefont {LG}~\bibnamefont {Leal}},\
  }\bibfield  {title} {\enquote {\bibinfo {title} {Migration of rigid spheres
  in a two-dimensional unidirectional shear flow of a second-order fluid},}\
  }\href@noop {} {\bibfield  {journal} {\bibinfo  {journal} {Journal of Fluid
  Mechanics}\ }\textbf {\bibinfo {volume} {76}},\ \bibinfo {pages} {783--799}
  (\bibinfo {year} {1976})}\BibitemShut {NoStop}%
\bibitem [{\citenamefont {Elfring}\ and\ \citenamefont
  {Lauga}(2015)}]{elfring2015theory}%
  \BibitemOpen
  \bibfield  {author} {\bibinfo {author} {\bibfnamefont {Gwynn~J}\ \bibnamefont
  {Elfring}}\ and\ \bibinfo {author} {\bibfnamefont {Eric}\ \bibnamefont
  {Lauga}},\ }\bibfield  {title} {\enquote {\bibinfo {title} {Theory of
  locomotion through complex fluids},}\ }in\ \href@noop {} {\emph {\bibinfo
  {booktitle} {Complex fluids in biological systems}}}\ (\bibinfo  {publisher}
  {Springer},\ \bibinfo {year} {2015})\ pp.\ \bibinfo {pages}
  {283--317}\BibitemShut {NoStop}%
\bibitem [{\citenamefont {Leal}(1975)}]{leal1975slow}%
  \BibitemOpen
  \bibfield  {author} {\bibinfo {author} {\bibfnamefont {LG}~\bibnamefont
  {Leal}},\ }\bibfield  {title} {\enquote {\bibinfo {title} {The slow motion of
  slender rod-like particles in a second-order fluid},}\ }\href@noop {}
  {\bibfield  {journal} {\bibinfo  {journal} {Journal of Fluid Mechanics}\
  }\textbf {\bibinfo {volume} {69}},\ \bibinfo {pages} {305--337} (\bibinfo
  {year} {1975})}\BibitemShut {NoStop}%
\bibitem [{Note2()}]{Note2}%
  \BibitemOpen
  \bibinfo {note} {A mathematica code for the evaluation of this surface
  integral is provided in the supplementary material.}\BibitemShut {Stop}%
\bibitem [{\citenamefont {Koch}\ and\ \citenamefont
  {Subramanian}(2006)}]{koch2006stress}%
  \BibitemOpen
  \bibfield  {author} {\bibinfo {author} {\bibfnamefont {Donald~L}\
  \bibnamefont {Koch}}\ and\ \bibinfo {author} {\bibfnamefont {G}~\bibnamefont
  {Subramanian}},\ }\bibfield  {title} {\enquote {\bibinfo {title} {The stress
  in a dilute suspension of spheres suspended in a second-order fluid subject
  to a linear velocity field},}\ }\href@noop {} {\bibfield  {journal} {\bibinfo
   {journal} {Journal of non-newtonian fluid mechanics}\ }\textbf {\bibinfo
  {volume} {138}},\ \bibinfo {pages} {87--97} (\bibinfo {year}
  {2006})}\BibitemShut {NoStop}%
\bibitem [{Note3()}]{Note3}%
  \BibitemOpen
  \bibinfo {note} {The coefficient is $ 6\pi \Psi _{1}^{*} $.}\BibitemShut
  {Stop}%
\end{thebibliography}%

\end{document}